\documentclass[twocolumn,showpacs,preprintnumbers,amsmath,amssymb]{revtex4}
\usepackage{tabularx,graphicx}\begin{document}
\newcommand{\beq}{\begin{equation}}
\newcommand{\eeq}{\end{equation}}
\newcommand{\beqn}{\begin{eqnarray}}
\newcommand{\eeqn}{\end{eqnarray}}
\newcommand{\bmath}{\begin{subequations}}
\newcommand{\emath}{\end{subequations}}

\title{Spin Meissner Effect in Superconductors and the Origin of the Meissner Effect}
\author{J. E. Hirsch }
\address{Department of Physics, University of California, San Diego\\
La Jolla, CA 92093-0319}


\begin{abstract}

We propose a dynamical explanation of the Meissner effect in superconductors and predict the existence of a spin Meissner effect:
that a macroscopic spin current flows within a London penetration depth $\lambda_L$ of the surface of superconductors in the absence of
applied external fields, with  carrier density =  the superfluid density and carrier speed $v=\hbar/(4m_e\lambda_L$) ($m_e=$bare
electron mass). The two members of a Cooper pair circulate in orbits of radius $2\lambda_L$ in opposite direction and
  the spin current in a Cooper pair has
orbital angular momentum $\hbar$. Our description also provides a 'geometric' interpretation of the difference between type I and type II superconductors. \end{abstract}
\maketitle 
 What is the $force$ that causes the electrons near the surface of a metal to start
moving and develop the electric current that will screen the magnetic field in the interior when the
metal is cooled into the superconducting state in the presence of an external static magnetic field?
Despite having been known for over 70 years, no $dynamical$ explanation of the Meissner effect exists
within the conventional understanding of superconductivity\cite{lorentz}.

Superfluid electrons carry one Bohr magneton of magnetic moment in addition to one electron charge. One may similarly
expect that the superfluid will respond to an $electric$ field by developing a $spin$ $current$\cite{sh1,sc}.
We assume there is no externally applied electric field, nevertheless  superfluid electrons interact with the electric field
originating in the positive ionic medium where they reside. 
Spin currents do not break time reversal invariance and hence should be insensitive to non-magnetic disorder.
Will superconducting electrons respond to the presence of $internal$ electric fields by developing a spin current?

In this paper we propose answers to both of these questions and in addition  we offer a new interpretation of what a Cooper pair is.
We rely on key physics proposed within the theory of hole
superconductivity\cite{holetheory}: that superconductivity is driven by kinetic energy lowering, that is achieved by 
$expansion$ of the electronic wave function\cite{hole2}, that also results in expulsion of negative charge from the interior towards
the surface\cite{expel}.

Consider a metal in a uniform magnetic field $B$ along the $\hat z$ direction. Its diamagnetic response (Landau diamagnetism) can be
understood as arising from induced Ampere circular currents of radius $r\sim k_F^{-1}$. For a nearly filled 
band (as required in the theory of hole superconductivity), 
$r\sim a$, the lattice spacing\cite{hole2}. 
Assume the transition to superconductivity involves a {\it radially outward} motion of   the  electrons in a Cooper pair {\it to a radius }
 $r=2\lambda_L$.  The electrons will acquire 
an azimuthal velocity $\vec{v}_\phi$ as they move outward 
{\it due to the magnetic Lorentz force} $\vec{F}_L=(e/c)\vec{v}\times\vec{B}$\cite{lorentz}. 
Conservation of angular momentum dictates\cite{rotating} that the azimuthal velocity at radius $r$ is  given by:
\beq
\vec{v}_\phi(\vec{r})=-\frac{e}{2m_e c}\vec{B}\times\vec{r}
\eeq
and for radius $r=2\lambda_L$
\beq
\vec{v}_\phi=-\frac{e\lambda_L}{m_e c}B\hat{\phi}.
\eeq

Note that the magnetic vector potential at position $\vec{r}$ {\it for a uniform magnetic field} is given by
$\vec{A}= \vec{B}\times\vec{r}/2$, hence Eq. (1) can be written as 
 \beq
\vec{v}_\phi=-\frac{e}{m_e c} \vec{A}
\eeq
In the conventional theory, Eq. (3) is understood as arising from the 'rigidity' of the 
superconducting wavefunction, which causes the canonical momentum $\vec{p}$ in the expression
\beq
m_e \vec{v}=\vec{p}-\frac{e}{c}\vec{A}
\eeq
to remain 0 independent of the value of the applied magnetic field\cite{tink}. Here instead, we derived this result from the $dynamics$.

When we now consider all the Cooper pairs in the superconductor, the 'internal' azimuthal velocities get cancelled by
superposition, as shown in Fig. 1(a), however those near the surface are not. 
The uncompensated azimuthal velocity Eq. (2) of the carriers  near the surface  
 is precisely what is needed to screen the magnetic field in the interior of the superconductor. For a long cylinder with  superfluid density $n_s$  the current density is
 \beq
\vec{j}_{charge}\equiv j = en_s\vec{v}_\phi.
\eeq
 Such current density flowing in a surface layer of thickness $\lambda_L$ induces a magnetic field in the interior of magnitude
\beq
B_{ind}=\frac{4\pi}{c}j\lambda_L=\frac{4\pi n_s e^2}{m_e c^2} \lambda_L^2 B
\eeq
and direction opposite to $\vec{B}$, that will exactly cancel $\vec{B}$ since\cite{tink}
\beq
\frac{1}{\lambda_L^2}=\frac{4\pi n_s e^2}{m_e c^2} .
\eeq

\begin{figure}
\resizebox{8.0cm}{!}{\includegraphics[width=7cm]{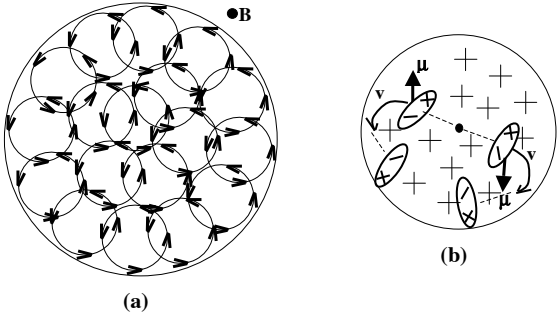}}
\caption{Illustration of the Meissner and spin Meissner effects. (a) shows the net direction of orbital motion of the 
{\it negatively charged} members of the Cooper pair
that expanded to radius $2\lambda_L$ (small circles in (a)), arising from the Lorentz force due to a magnetic field pointing out of the paper. Note how the 'internal' velocities
get cancelled and there is a net counterclockwise motion near the surface. (b) shows the spin Meissner effect for one of the circles in (a): up and down magnetic moments
get deflected in opposite direction due to the torque exerted by the
radially pointing electric field from the positive charge distribution. (b) also shows the equivalent electric dipoles as they move
out. Internal dipoles  cancel out and only the dipoles near the surface of the sample survive. }
\label{atom5}
\end{figure}

Having thus provided a 'dynamical' explanation of the origin of the ordinary Meissner effect, we now turn our attention to the spin Meissner effect. 
We choose the $\hat{z}$ axis
as the spin quantization axis, hence the electron magnetic moment is 
\beq
\vec{\mu}=\mu_B\sigma\hat{z}=\frac{e\hbar}{2m_ec}\sigma \hat{z}
\eeq
with $\sigma=+/-1$. 
Consider now the effect of a uniform positive background charge density on a radially outgoing magnetic moment $\vec{\mu}$ (Fig. 1 (b)). The moving magnetic moment is 
equivalent to an electric dipole $\vec{p}$ in the laboratory frame\cite{boyer,dipole}
\beq
\vec{p}=\gamma \frac{\vec{v}}{c}\times\vec{\mu}
\eeq
($\gamma\sim1$) on which a radial electric field $\vec{E}_i$ will exert a torque $\vec{\tau}=\vec{p}\times\vec{E}_i$, causing a change in orbital angular momentum 
\beq
\frac{d\vec{L}}{dt}=m_e\frac{d}{dt}(\vec{r}\times{\vec{v}})=(\frac{\vec{v}}{c}\times\mu)\times\vec{E}_i
\eeq
The electric field of a cylindrical uniform charge density $\rho$ is $\vec{E}_i=2\pi\rho  \vec{r}$. 
The positive ionic charge density that the superfluid senses is $\rho=|e|n_s$. Hence from Eq. (10)
\beq
m_e\vec{r}\times\frac{d\vec{v}}{dt}=2\pi e n_s\vec{r}\times (\frac{\vec{v}}{c}\times\vec{\mu}) 
\eeq
Eq. (11) can be interpreted as if an effective spatially uniform magnetic field\cite{dipole,chudnovsky}
 
\beq
\vec{B}_\sigma= 2\pi n_s\vec{\mu}
\eeq
is exerting a Lorentz force $\vec{F}=(e/c)\vec{v}\times\vec{B}_\sigma$ on the electron. Just as in the earlier discussion, it will give rise to an azimuthal velocity
\beq
\vec{v}_{0\sigma}=-\frac{e\lambda_L}{m_e c}B_\sigma\hat{\phi}=-\frac{2\pi e n_s \lambda_L}{m_e c}\vec{\mu}\times \hat{r}
\eeq
for outward motion to radius $r=2\lambda_L$, where now the direction of the
azimuthal velocity is opposite for up and down spin electrons.
Again, the 'internal' azimuthal velocities get cancelled by superposition and the uncompensated azimuthal velocities 
  give rise
to a macroscopic spin current
\beq
\vec{j}_{spin}\equiv\frac{n_s}{2}(\vec{v}_{0 \uparrow}-\vec{v}_{0 \downarrow}) =n_s \vec{v}_{0\uparrow}
\eeq
flowing within a London penetration depth of the surface, whether or not an external magnetic field is applied.

Just like in the case of the real magnetic field, the effective magnetic field Eq. (12) can be understood as arising from a vector potential
\beq
\vec{A}_\sigma=\frac{\vec{B}_\sigma\times\vec{r}}{2}=2 \pi n_s \lambda_L \vec{\mu}\times\hat{r}
\eeq
and the total velocity 
 \beq
\vec{v}_\sigma=-\frac{e}{m_ec}\vec{A}-\frac{e}{m_ec}\vec{A}_\sigma =  \vec{v}_\phi +\vec{v}_{0\sigma}
\eeq
can be interpreted as arising from the 'rigidity' of the wave function as in Eq. (4).

Note that the  Dirac Hamiltonian for an electron including the spin-orbit interaction  in the non-relativistic limit in the presence of an electric field $\vec{E}$  yields\cite{bd}
\beq
H=\frac{1}{2m_e}  (\vec{p}-\frac{e}{c}\vec{A})^2 -\frac{1}{2m_e c}\vec{p} \cdot (\vec{\mu}\times\vec{E})
\eeq
The last term in Eq. (17) has been interpreted
 as a magnetic vector potential giving rise to an 'effective' magnetic field.\cite{ac,chudnovsky} 
However, with $\vec{E}=\vec{E}_i$ it yields an effective magnetic field {\it of opposite sign} to Eq. (12), so this argument  is misleading. That the force on an electron with  magnetic moment $\vec{\mu}$  moving with velocity $\vec{v}$ in a $positive$ background is in  direction $e\vec{v}\times\vec{\mu}$, as predicted by Eq. (12), also follows from the treatment of  Ref. \cite{dipole}.

Using Eq. (7), the spin current velocity Eq. (13) can be written in the remarkably simple form
\beq
\vec{v}_{0\sigma}  =-\frac{\hbar}{4 m_e \lambda_L}\vec{\sigma}\times \hat{r}
\eeq
which is the central result of this paper. 

How large is this azimuthal velocity? For a typical $\lambda_L=400\AA$, Eq. (18) yields $v_{0\sigma}=72,395 cm/s$. The kinetic energy per carrier associated with it is
\beq
\frac{1}{2} m_e v_{0\sigma}^2=\frac{0.238}{\lambda_L(\AA))^2} eV=\frac{\pi}{2} n_s \mu_B^2
\eeq
which yields $1.49 \mu eV$ for $\lambda_L=400\AA$, which is comparable to superconducting condensation energies per electron. Since each superfluid electron in the
bulk lowers its energy by the condensation energy, there is plenty of energy available to account for the kinetic energy of the spin current that resides only in the
surface layer of thickness $\lambda_L$. Note also the form of the last term in Eq. (18) (obtained using Eq. (7)): 
it says that for two oppositely alligned spin 1/2 magnetic moments perpendicular to the
radius joining them at distance of order the interelectron distance ($n_s^{-3}$), the energy lowering from magnetic dipole-dipole
interaction can account for the spin current kinetic energy cost. 

There is another independent argument that supports the validity of Eq. (18). An applied magnetic field $\vec{B}=B\hat{z}$ will 
exert 
a Lorentz force on the superfluid carriers in the spin current in the surface layer. This force
\beq
\vec{F}_{L,\sigma}=\frac{e}{c}\vec{v}_{0\sigma}\times\vec{B}=-\frac{e v_{0\sigma}}{c}\sigma B \hat{r}
\eeq
points $inward$ for magnetic moment parallel to $\vec{B}$ ($\sigma=-1$) and outward for opposite spin. 
In addition, because $B$ only penetrates a distance $\sim 2\lambda_L$\cite{note}, 
its gradient exerts a force on the magnetic moments that is $outward$ for magnetic moment parallel to $B$ 
 and inward for opposite spin:
\beq
\vec{F}_{g,\sigma}=\vec{\nabla}(\vec{\mu} \cdot \vec{B})=
\frac{e\hbar}{2m_ec}\sigma \frac{\partial B}{\partial r}\hat{r}
\sim \frac{e\hbar}{4m_e c\lambda_L}\sigma  B\hat{r}
\eeq
where we have used $\partial B /\partial r \sim B/\ 2 \lambda_L$.
If we require that the forces Eq. (20) and (21) exactly cancel for each spin component  we obtain the value Eq. (18) for the spin current velocity.
Hence an applied 
$B$ will not affect the radial distribution of the spin current. Note how in this argument the orbital and spin aspects are inextricably intertwined.

What is the magnitude of a real magnetic field that would give rise to the same azimuthal speed of the superfluid as the speed of the spin current? Equating Eqs. (2) and (18) it follows it is simply $|\vec{B}_\sigma|$ (Eq. (12)), which can also be written as 
\beq
B_s \equiv |\vec{B}_\sigma|=-\frac{\hbar c}{4e\lambda_L^2}=\frac{\phi_0}{4\pi \lambda_L^2}
\eeq
with $\phi_0=hc/2|e|$ the flux quantum. Application of an external magnetic field of magnitude $B_s$ would cause one of the components of the spin current to 
move twice as fast and the other one to stop altogether.
However, note that this magnetic field is essentially the lower critical field of a type II superconductor\cite{tink}
\beq
H_{c1}\sim \frac{\phi_0}{4\pi \lambda_L^2} ln\kappa
\eeq
with $\kappa$ the Ginzburg Landau parameter ($\kappa>1/\sqrt{2}$ for type II superconductors). Hence, when a 
magnetic field is applied to a type II superconductor large enough that it causes one of the spin current components to
 essentially stop ($B=B_s\sim H_{c1}$), superconductivity is destroyed and the magnetic field penetrates the sample forming flux tubes of normal 
material threaded by magnetic field $H_{c1}$.

What about type I superconductors? The thermodynamic critical field for a free electron model can be written as\cite{tink}
\beq
H_c=\sqrt{\frac{3}{2}}\frac{1}{\pi^2}\frac{\phi_0}{\xi_0 \lambda_L}
\eeq
with $\xi_0$ the Pippard coherence length. Hence from Eqs. (22) and (24), $B_s=0.64(\xi_0/\lambda_L)H_c$.
 Since for a type I superconductor $\xi_0>\sqrt{2}\lambda_L$, the superconducting state will be destroyed well before the applied magnetic field reaches $B_s$.
  \begin{figure}
\resizebox{8.5cm}{!}{\includegraphics[width=7cm]{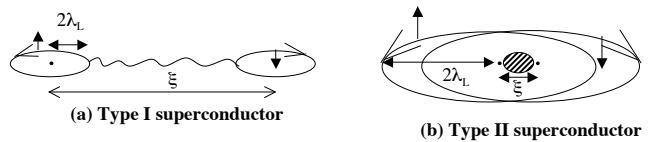}}
\caption{ In type II materials ((b)) with $\xi<2\lambda_L$, a normal vortex core of diameter $\xi$ can be enclosed by both orbits of the same Cooper pair.
In type I materials ((a)), $\xi>2\lambda_L$ and  this cannot happen. }
\label{atom6}
\end{figure}

What is the role of the coherence length $\xi$? We propose that 
{\it  $\xi$ is the average distance between the centers of the orbits of the two members of a Cooper pair}, as shown schematically in Fig. 2.
This is consistent with the conventional understanding that $\xi$ is the 'size' of the Cooper pair. However, in our case it provides a new {\it geometric
interpretation} for why the  cross-over between type I and type II behavior occurs for $\xi \sim 2\lambda_L$ . This will be discussed in more
detail in a forthcoming paper.

How can one detect  the existence of this spin current in superconductors experimentally? (i) Inserting in a superconducting ring a spintronics device that rectifies
spin current\cite{spintronics} one would measure an unexpected asymmetry depending on its orientation; 
(ii) Spin-resolved  neutron scattering with very cold neutrons may be able to detect it; 
(iii) Since the spin current gives rise to equivalent electric dipoles, a superconducting ring 
will have electric field lines as shown in Fig. 3, with a non-zero quadrupole moment which we can estimate as
\beq
Q=-4\pi a^2 h \lambda_L n_s p
\eeq
with $a$ the radius, $h$ the height, $n_s$ the superfluid density and $p$ the electric dipole moment Eq. (9).  For $\lambda_L=400\AA$,
$p=2.24\times 10^{-26} esu$, $n_s=1.77\times10^{22}/cm^3$ and
\beq
Q=-1.99\times 10^{-8} a(cm)^2 h(cm) esu
\eeq
One may attempt to measure the electric field around the ring or the electrostatic
force between two such rings. For two rings on top of each other (Fig. 4a) a distance $z$ apart the force is repulsive
and given by $F_z=5Q^2/z^6$ and should be detectable for small enough $z$ (note that $z\geq h)$. For different relative position and orientation of the rings the force will be repulsive or attractive as shown in Fig. 4.

\begin{figure}
\resizebox{4.5cm}{!}{\includegraphics[width=7cm]{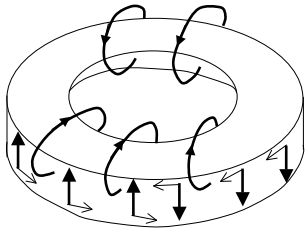}}
\caption{Pattern of electric field (curved lines) expected to arise for a superconducting ring. The vertical arrows denote the direction of the magnetic moment of the superelectrons and
the horizontal arrows the direction of their motion. No magnetic field nor charge current is present.}
\label{atom5}
\end{figure}

 \begin{figure}
\resizebox{5.5cm}{!}{\includegraphics[width=7cm]{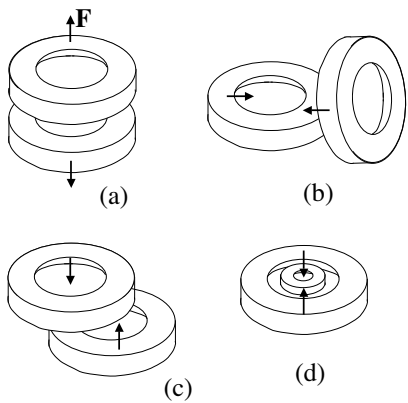}}
\caption{Force between two superconducting rings due to the electric fields produced by the spin currents.
Depending on their relative position and orientation the force can be repulsive (a) or attractive (b), (c). In (d), a smaller ring should be in stable equilibrium at
the center of a larger ring.}
\label{atom6}
\end{figure}

In a spherical or planar geometry in the absence of external fields there is no preferred direction to the spin current.
However applying a  small magnetic field will define a spin quantization direction and orient the spin current flow
according to Eq. (14).
An externally applied electric field is  expected to modify the spin current at
 low temperatures. We point out that the magnitude of the spin current density $e\vec{j}_{spin}$  predicted by this physics can be as high as
$10^8 Amps/cm^2$ and  may provide a useful source of spin current for spintronics applications.

 Why do superfluid carriers near the surface 'choose'
 to pay the kinetic energy Eq. (19) in the absence of externally applied fields, 
rather than stop and save that energy altogether? Those carriers are tasked with shielding the interior of the superconductor from
applied magnetic fields by setting up a Meissner current. Thus we may think of them as
 'confined'   in the surface layer of thickness $\sim 2\lambda_L$, ready to respond.
 The quantum zero-point motion energy of  an electron confined in a region of linear dimension $\Delta x$ is $\hbar^2/2m_e (\Delta x)^2$,
 of the same order as the kinetic energy Eq. (19). 
 
 Why is the radius of the orbit $2\lambda_L$?  $2\lambda_L=\sqrt{\hbar c/|e B_\sigma|}$, the spin-orbit `magnetic length'. The flux of $B_\sigma$  through the circle of radius $2\lambda_L$ is
 \beq
 \phi=\int \vec{B}_\sigma \cdot d\vec{A} =\oint \vec{A}_\sigma \cdot d\vec{l}=\frac{hc}{2e}=\phi_0,
 \eeq
 the flux quantum. The spin current velocity Eq. (18) corresponds to an orbital angular momentum 
 $m_e v _{0\sigma} r=\hbar/2$ for an electron 
 in a circular orbit of radius $r=2\lambda_L$. 
 As the Cooper pair is born, its wave function expands so that the orbit of each member of the Cooper pair 
  will thread one  flux quantum of spin-orbit interaction flux,
 and in the process each member of the Cooper pair acquires orbital angular momentum $\hbar /2$, equal to its spin angular momentum.
 Thus  the Cooper pair wave function is single-valued, and the factor of $2$ in $\phi_0=hc/2e$ acquires a new meaning.
 
\acknowledgements
The author is grateful to Congjun Wu for helpful discussions.
  

\begin{references}
\bibitem{note} Note that a magnetic field $B$ decreasing linearly to zero over a distance $2\lambda_L$ from the surface corresponds to a
vector potential $A=B\lambda_L$ as required for the Meissner effect.
\end{references}
 \end{document}